\documentstyle[11pt,newpasp,twoside,epsf]{article}
\begin{document}
\title{The Crab glitches: incidence and cumulative effect}
 \author{F.Graham Smith \& C. Jordan}
\affil{Jodrell Bank Observatory,
University of Manchester, Macclesfield, Cheshire SK11 9DL, U.K.}

\begin{abstract}
The fourteen glitches observed during 33 years do not show
the simple pattern expected from a relaxation oscillator.  They may
however be regarded as three major events separated by about 12 years,
the third being a group of smaller glitches.  There is a step increase
in slowdown rate at each glitch, whose cumulative effect makes a
significant contribution to the second differential $\ddot\nu$.  The
braking index $n$ has previously been evaluated only between glitches:
the effect of the glitches is to reduce $n$ from 2.51 to 2.45.  This
extra effect due to the glitches would be explained by an increase in
dipole field at the fractional rate of $1.5\times10^{-5}$ per annum.
\end{abstract}

\section{Introduction}
I start by reminding you that we do not fully understand the
rotational slowdown of pulsars.

The rotational slowdown of neutron star losing angular momentum by
pure magnetic dipolar radiation should follow a power law
\begin{equation} \dot\nu\propto -\nu^n \end{equation} where the
braking index $n=3$. This can be tested when a value of $\ddot \nu$
can be measured, by finding \begin{equation} n=\frac{\nu
\ddot\nu}{\dot \nu^2} \end{equation} Where measurements are possible,
no pulsar has been found with a braking index of 3.  The discrepancy
is usually attributed to complications in the magnetosphere, and
particularly to angular momentum loss by particle outflow in the polar
regions.

The slowdown is punctuated by glitches. Measurements of $\ddot \nu$
require long runs of data, which for most pulsars include glitches; we
do not know how these affect the long-term behaviour.  Taking the view
that glitches are temporary steps followed by a complete recovery, we
can smooth over several glitches to find $\ddot\nu$; for the Vela
pulsar this gives $n=1.4\pm0.2$.  For the Crab pulsar the glitches are
small and there are sufficiently long stretches of data between them
to allow measurement of $\ddot\nu$, giving remarkably consistent
values of braking index around $n=2.51$.  But we know that for the
Crab and possibly other pulsars there is a persistent step in slowdown
rate $\dot\nu$ at glitches.  What is the effect of these steps on the
overall slowdown law?

We now have available an almost complete record of the rotational
slowdown of the Crab pulsar over 33 years (Lyne et al 1993, Wong et al
2001), including 14 glitches above a well-defined size limit.  We
expect to publish details of the most recent glitches shortly (Jordan,
Smith and Lyne in preparation).  These Crab pulsar glitches are
comparatively small and infrequent, offering the possibility of
comparing their effect on the overall long-term slowdown with the
smooth behaviour between glitches.

\section{Incidence of the glitches}

\begin{figure}
\plotone{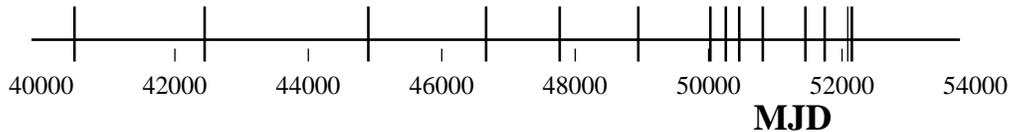}
\caption{The sequence of glitches in the Crab pulsar 1969-2002.}
\label{occur}
\end{figure}

The Julian dates of occurrence of the glitches are shown in Fig \ref{occur}.
Can we characterise this sequence in terms of a simple stick-slip
behaviour of a two-component system, comprising the solid crust and
part of the superfluid?  We would expect to see a relaxation
oscillator, with a more regular sequence.  Must we consider a
multicomponent system which would produce some sort of random or
chaotic sequence of occurrence and size of glitches?  If we consider
only the intervals between glitches, without regard to their sequence,
we do not see anything like a relaxation oscillator; in contrast the
intervals between glitches are spread in a reasonable approximation to
a Poisson distribution.  But the actual sequence does not seem to be
random.

\begin{figure}
\plotone{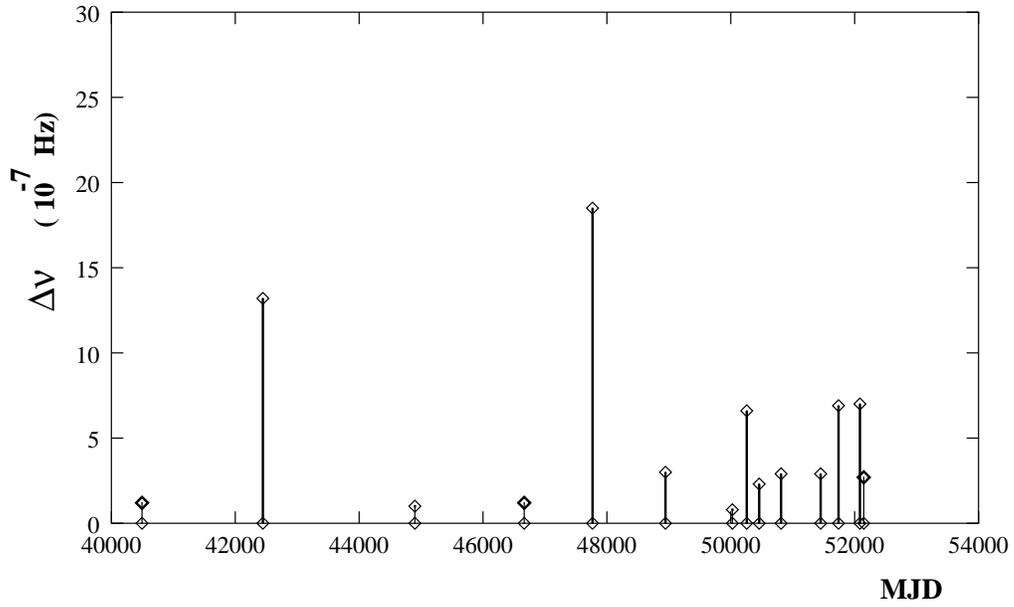}
\caption{The size of each glitch measured by the height of the initial step.}
\label{height}
\end{figure}

Fig \ref{height} shows the size of each glitch as measured by the
initial step $\Delta\nu$.  A possible description is now in terms of
three events, the first two being the large glitches in 1975 and 1989,
and the third being a closer spaced group of smaller glitches.  These
major events were spaced by about 12 years, so it will take some time
before we can confirm this interpretation.

\begin{figure}
\plotone{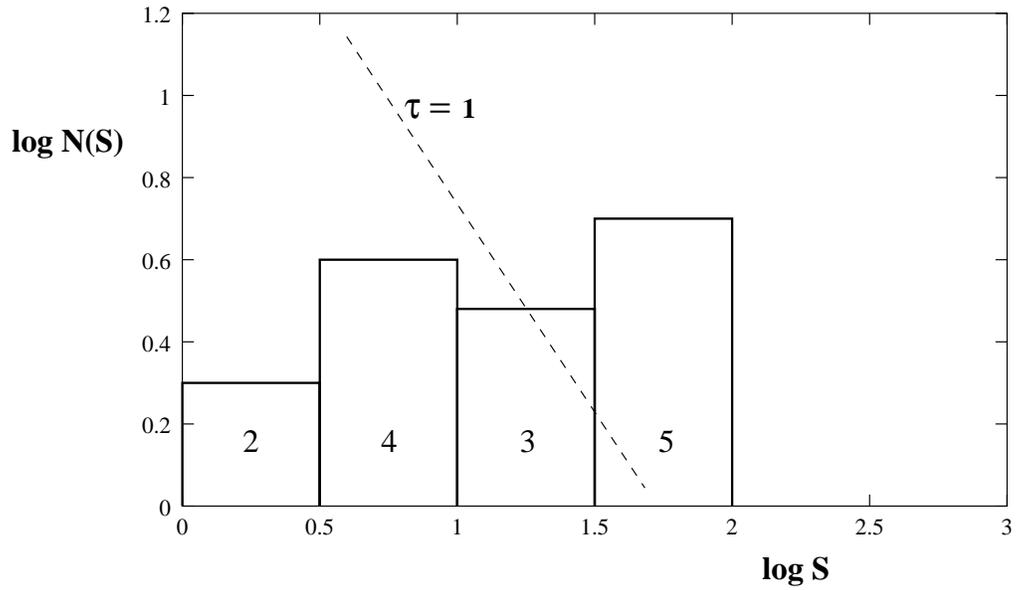}
\caption{The logarithmic size/number relation for Crab pulsar glitches.}
\label{logN/S}
\end{figure} 

Is it reasonable to consider the Crab glitches as characteristic of a
simple two-component system?  The more random behaviour of the Crab
pulsar might be indicating a many-component system, typified by
earthquakes or the classical sandpile.  In such systems the frequency
of events $N$ is related to event size $S$ by a power law $N\propto
S^{-\tau}$, where $\tau$ is commonly in the range 1 to 2 (Per Bak
1996).  Figure \ref{logN/S} shows the statistics of glitch size on a
log plot, with a line corresponding to $\tau=1$.  There is an obvious
lack of both small and large events for this type of behaviour.  We
conclude that the system is basically a two-element system, with some
variation in the triggering which led to the group of smaller
glitches.

\section{The cumulative effect on slowdown rate}

\begin{figure}
\plotone{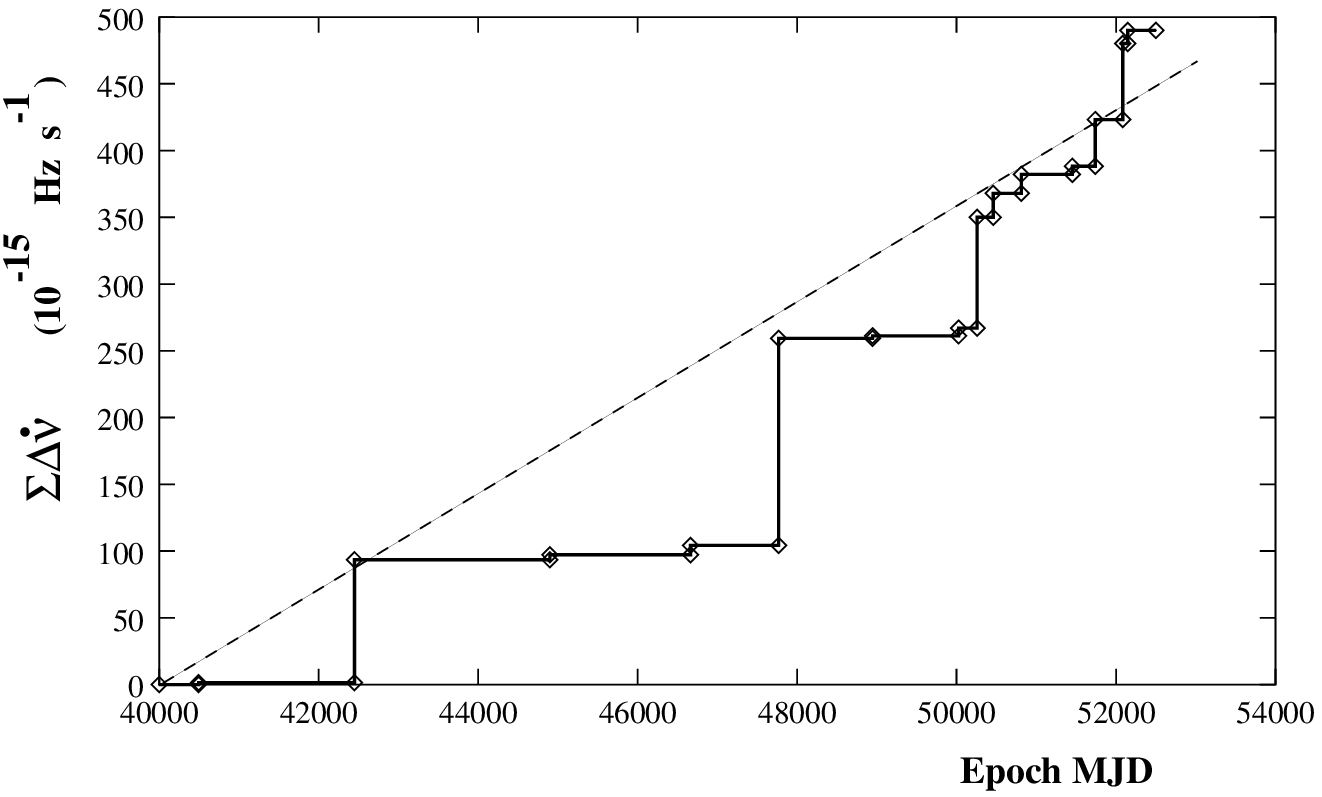}
\caption{The cumulative effect of the steps $\Delta\nu$ in slowdown rate.}
\label{cumulative}
\end{figure} 

We turn now to the steps $\Delta\dot\nu$ in slowdown rate at the
glitches.  Fig \ref{cumulative} shows the cumulative effect of the
steps in slowdown rate at the glitches.  This provides a significant
contribution $\ddot\nu_{\rm g}$ to the overall second differential
$\ddot\nu$; we identify this glitch contribution as the overall slope of Fig
\ref{cumulative}.  The line arbitrarily drawn has a slope
corresponding to $\ddot\nu_{\rm g}= -6.2\times10^{-23}$ Hz s$^{-2}$.

The derivative of the slowdown rate has previously been evaluated in
the smooth runs between widely separated glitches, giving
$\ddot\nu_{\rm g}= 1.185\times10^{-20}$ Hz s$^{-2}$; this led to the
widely quoted value $n=2.51$ for the braking index.  The contribution
$\ddot\nu_{\rm g}$ is a 3\%
 {\it reduction} in $\ddot\nu$, giving a new
value $n=2.45$ for the long-term slowdown.

\begin{figure} 
\plotfiddle{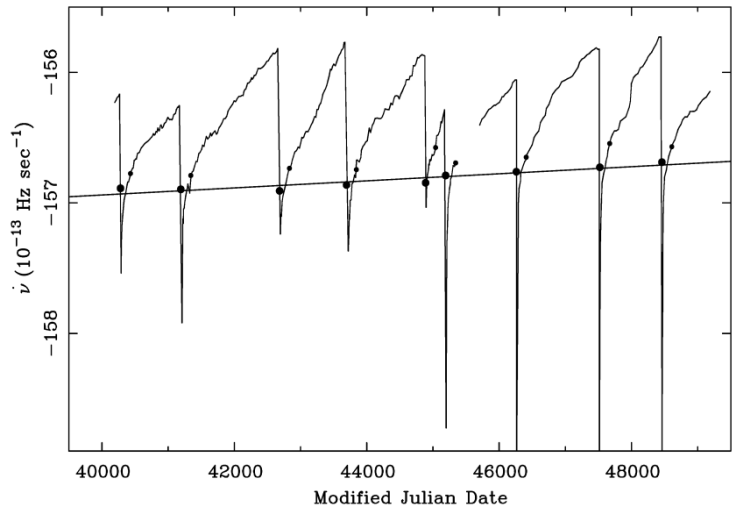}{200pt}{0}{33}{33}{-144pt}{0pt}
\caption{The slowdown rate of the Vela pulsar over 25
years (Lyne et al. 1996), showing the slow change in the second
differential $\ddot\nu$.}  
\label{vela}
\end{figure}

This new value can be compared with the braking index of the Vela
pulsar, which can only be evaluated by integrating the effect of many
glitches (Fig \ref{vela}, Lyne et al. (1996) to produce a value of
$\ddot\nu$. This is the origin of the measured value of the index
$n=1.4\pm0.2$.

\section{The dynamics of slowdown}

We return to the question: why is the braking index different from 3?
Following Allen and Horvath (1997), there are evidently two separate
contributions, which we call interior and exterior, due respectively to glitches
and to the magnetosphere. The magnetospheric contribution is due to
the overall configuration of the magnetic field, including the effect
of angular momentum loss by particle outflow.  The glitches provide
the interior contribution, which we expect to be a relaxation
oscillation with no average effect.  But in fact they produce an
accumulating effect on slowdown rate; how does this happen?  Are any
of the parameters in the dipole slowdown law Equation \ref{law}
changing at each glitch?

\begin{equation} \dot\nu=-M^2 \sin^2\alpha I^{-1}\nu^3 \label{law}
\end{equation} where $M$ is the dipole moment, $\alpha$ is the
inclination of the dipole to the rotation axis and $I$ is the moment
of inertia.

Is $I$ changing?  For the Crab the glitches increase the slowdown rate
by a fraction $0.34\times 10^{-4}$ per year.  This is too large to be
due to a reduction in ellipticity, since the equilibrium value of
ellipticity is $e=10^{-4}$.

An apparent reduction in $I$ might be due to a continuous accumulation
of pinned vortices, in `capacitors' (Alpar et al. 1996), eventually
locking up a large fraction of the superfluid.  This would have to
persist and accumulate through a succession of catastrophic glitches,
which seems unlikely.  We therefore look at a change in $M_\perp$,
either in the dipole moment or in the misalignment angle $\alpha$.

Is $\alpha$ changing? The slowdown rate is proportional to
$\sin^2\alpha$, so that the required fractional change in $\alpha$ is
greater by ${1\over 2}\tan\alpha$ than $\ddot\nu_{\rm g}/\ddot\nu$.
Romani \& Yadigaroglu (1995) show that $\alpha\approx 70^\circ$,
giving a required increase of about $10^{-4}$ radians per year.
According to their model the spacing between the two
main pulse components is sensitive to such a change; putting
$\alpha=70^\circ$ we find the spacing would increase by about 2
microseconds per year.  This probably allows a check to be made, but
in any case the distribution of $\alpha$ amongst older pulsars does
not support the idea of evolution towards orthogonality.

So we arrive again at the proposal for an increasing dipole field,
which has already been suggested for the Vela pulsar (Lyne, Shemar \&
Smith 2000). For the Crab pulsar the fractional increase would be
$1.5\times 10^{-5}$ per annum.  Considering that we know very little
about the generation of the dipole field, there seems to be no argument
against this suggestion in a young pulsar like the Crab.

Accounting for the step changes in slowdown at glitches in this way
does not, of course explain the major part of the deviation of the
braking index from the theoretical value $n=3$.  Glitches are a
phenomenon of the neutron star; the main anomaly in the Crab pulsar is
evidently concerned with the configuration of the magnetosphere.  We
do not know if this true also of the Vela and other pulsars.


\begin{references}

Allen, M. P.\& Horvath, J. E. 1997, \apj, 488, 409\\
Alpar, M. A., Chau, H. F., Cheng, K. S. \& Pines, D., 1996, \apj, 459, 706\\
Lyne, A. G., Pritchard, R. S. \& Smith, F. G. 1993 \mnras, 265, 1003\\  
Lyne, A. G., Pritchard, R. S., Smith, F. G. \& Camilo, F. 1996, Nature, 381, 497\\  
Lyne, A. G., Shemar, S. L. \& Smith, F.,G. 2000, \mnras, 315, 534\\
Per Bak 1996, in How Nature Works, Springer-Verlag, New York\\
Romani, R. W. \& Yadigaroglu, L-A. 1995, \apj 438, 314\\
Wong, T., Backer, D. C. \& Lyne, A. G. 2001 \apj, 548, 447\\
\end{references}
\end{document}